\documentclass[pra,showpacs,twocolumn,nofootinbib]{revtex4}
\usepackage{graphicx}
\setkeys{Gin}{draft=false}
\usepackage{bm,amssymb,amsmath}
\usepackage{dcolumn}
\usepackage{natbib}
\usepackage{graphicx}

\newcommand{\fref}[1]{Fig.~\ref{#1}}

\newcommand{\Eref}[1]{Eq.~(\ref{#1})}
\newcommand{\tref}[1]{Table~\ref{#1}}

\newcommand{\cm}{cm$^{-1}$}

\begin{document}

\title{Sensitivity coefficients to $\alpha$-variation for astrophysically
relevant transitions in Ni II}
\author{E. A. Konovalova$^{1}$}
\author{M. G. Kozlov$^{1,2}$}
\author{R. T. Imanbaeva$^{2,1}$}

\affiliation{$^1$Petersburg Nuclear Physics Institute, Gatchina 188300,
Russia}

\affiliation{$^2$St.~Petersburg Electrotechnical University ``LETI'', Prof.
Popov Str. 5, 197376 St.~Petersburg}

\date{
\today}

\begin{abstract}
We calculated the dependence of the transition frequencies on the
fine-structure constant $\alpha$ ($q$-factors) for Ni~II. Nickel is one of the
few elements with high sensitivity to $\alpha$-variation, whose lines are
observed at high redshifts. This makes it a sensitive probe for
$\alpha$-variation on the cosmological timescale. The electronic structure of
Ni II ion was treated within the configuration interaction (CI) method using
Dirac-Coulomb Hamiltonian.
\end{abstract}
\pacs {31.15.aj, 06.20.Jr, 31.15.am, 23.20.Lv}

\maketitle
\section{Introduction}

There are two dimensionless constants, the fine-structure constant ($\alpha$)
and the proton-electron mass ratio ($\mu$), for which spectroscopy is a test
ground to probe temporal and spatial variations. The constant $\alpha$ is
important for electronic structure of atoms and molecules. The dependence of
the spectrum on $\alpha$ appears through the relativistic corrections, such as
the fine-structure, Lamb shift, etc. The leading relativistic corrections
scale as $\alpha^2Z^2$ where $Z$ is atomic number. Therefore, heavier elements
have higher sensitivity to $\alpha$-variation. For the astrophysically
relevant elements $Z<30$ and relativistic corrections do not exceed few
percent. The constant $\mu$ enters atomic spectra only through the isotope
shifts which are typically on the order of $10^{-3}$, or less. Due to the
presence of the vibrational and rotational structures, the dependence of the
molecular spectra on $\mu$ is much more pronounced. Therefore, the search for
$\alpha$-variation is usually done using atomic spectra
\cite{DFW99a,DFW99b,WFC99,LML07,WKM10,MCWEM13,SoCo14}, while $\mu$-variation
is studied with the help of molecular spectra
\cite{Tho75,KMU11,WM12,RWSN13,ARNP14,BUMW14}. Let us note in passing that much
higher sensitivity to the variation of both fundamental constants than in
optics, can be found in the infrared and microwave wavebands
\cite{LCB12,KoLe13}.

Astronomical differential measurements of the constant $\alpha$ is based on
the comparison of the line centers in the absorption or emission spectra of
cosmic objects and the corresponding laboratory values. It follows that the
uncertainties of the laboratory rest frequencies and of the line centers in
astronomical spectra are the prime concern of such measurements. For example,
the unknown isotopic abundances of the elements in the Universe can lead to
the systematic frequency shifts, comparable to the expected signal from
$\alpha$-variation \cite{KKBD04}. That is why it is important to use such
heavy elements as Fe and Ni where relativistic effects are larger, while
isotopic shifts are suppressed.

Most studies of the possible $\alpha$-variation at the redshifts $z\gtrsim 1$
are based on the analysis of the quasar absorbtion spectra using many
multiplet method suggested in Refs.\ \cite{DFW99a,DFW99b,WFC99}. This method
utilizes lines of different ions from the same source. All these lines are
analyzed simultaneously to find the redshift and the value of $\alpha$ for
this object. The lines with small sensitivity coefficients to
$\alpha$-variation serve as anchors which give the redshift, while the lines
with high sensitivity serve as probes, which give the value of $\alpha$. All
lines of the light ions, like Mg~II and Al~II, belong to the first category.
Most lines of heavier ions, like Fe~II, Ni~II, and Zn~II, fall into second
category. However, some of the lines of Ni~II have relatively small
sensitivity and belong to anchors. This can be important for the control of
the systematics. The lines of the ion Fe~II have sensitivity coefficients to
$\alpha$-variation of different signs. This also allows effective control of
the possible systematic effects.

At present there is one group, that reports nonzero space-time variation of
$\alpha$ on the level of few parts per million (ppm) \cite{WKM10} (known as
the ``Australian dipole''). These results have not been confirmed by other
groups, who give only upper limits on the $\alpha$-variation also at the ppm
level (see \cite{LML07,RSGP12,MCWEM13,SoCo14} and references therein). Because
of that it is important to continue observations and include more sources and
additional lines. Such programs are currently going on at the VLT and Keck
telescopes \cite{MCWEM13,SoCo14}.

A list  of the astrophysically relevant optical lines of atoms and ions is
given in compilations \cite{BDFK10,MuBe13}. These compilations includes eleven
lines of Ni~II. According to \citet{HS13} there are nine lines of Ni~II, which
can be observed in the quasar spectra including one line not listed in
\cite{BDFK10,MuBe13}. The rest frequencies for Ni II are tabulated in NIST
database \cite{NIST}. Compilation by \citet{Mor03} includes both frequencies
and oscillator strengths $f_\mathrm{osc}$ for astrophysically relevant lines.
Several high accuracy laboratory rest frequencies were measured in Ref.\
\cite{PTML00}. According to \cite{MuBe13} more accurate laboratory
measurements are needed. The sensitivity to $\alpha$-variation ($q$-factors)
was theoretically studied in Refs.\ \cite{MWFD01,DFK02} for four lines from
the list \cite{HS13}. In this paper we do calculations of the $q$-factors and
oscillator strengths $f_\mathrm{osc}$ for all nine astrophysically interesting
lines of Ni~II.

\section{Theory and Method}

Supposing that the nowadays value of $\alpha$ differs from its value in the
earlier Universe we can study space-time variation of $\alpha$ by comparing
atomic frequencies for distant objects in the Universe with their laboratory
values. In practice, we need to find relativistic frequency shifts, known as
$q$-factors \cite{DFW99b,DFW99a,DFMW01,DFM03}. The difference between the
transition frequencies in astrophysical spectra and the laboratory ones is
given by the formula:
\begin{equation}
\begin{aligned}
\label{eq:theory1}
 \omega_{i} = & \omega_{i}^{(0)} +  \omega_{i}^{(2)}\alpha^{2} + \ldots
 = \omega_{i,\mathrm{lab}} + q_{i}x + \ldots , \\
 & \omega_{i,\mathrm{lab}}=\omega_{i}^{(0)} +  \omega_{i}^{(2)}\alpha_{0}^{2}, \\
 & x=(\alpha/\alpha_{0})^2 - 1,  \quad q=\partial \omega / \partial x\Bigr|_{x=0}
 ,
\end{aligned}
\end{equation}
where $\omega^{(0)}$ is a transition frequency in non-relativistic
approximation, $\omega^{(2)}$ is a relativistic correction. In Eq.~(\ref{eq:theory1}) $\omega_{i,\mathrm{lab}}$ is the frequency value for
$\alpha = \alpha_{0}$, where $\alpha_{0}$ is fine-structure constant in the
laboratory (present time) conditions. Note that transition $q$-factor is
simply a difference between $q$-factors of upper and lower levels.

For the search of the $\alpha$-variation it is most advantageous to use atoms
and ions for which $q$-factors of transitions between certain states
significantly differ from each other. That means we need elements with high
$Z$. At the same time these elements should be abundant in the Universe to
provide sufficient observational data. The latter restriction leaves us with
Fe and Ni as the most heavy abundant elements. Fe has been studied in detail
in Refs.\ \cite{PKT07, DF08a,PKR09} and other relevant elements have been
studied in \cite{BDFM04, BFK05, BFK06, DJ07, SD08}. Ni has been investigated
in a lesser detail \cite{DFK02, MWFD01} and not all levels observed in
astrophysics have been calculated. This is why we return to this problem here.

Rough estimates of the $q$-factors can be obtained from a simple one-particle
model \cite{DFW99a}. But in order to obtain more accurate values one has to
account for electronic correlations and perform large-scale numerical
calculations. To find $q$-factors numerically we need to solve the atomic
relativistic eigenvalue problem for different values of $\alpha$, or
equivalently for different values of $x$ from \Eref{eq:theory1}. In this case
we can get $q$-factor as:
\begin{equation}
\begin{aligned}
\label{eq:theory2}
q \approx \frac{\omega(x_{+})-\omega(x_{-})}{x_{+}-x_{-}}\, .
\end{aligned}
\end{equation}
Our previous experience shows that convenient choice is $x_\pm=\pm \tfrac18$.
In order to test the accuracy of this approximation we can also estimate the
second derivative:
\begin{equation}
\label{eq:theory3}
 \left.\frac {\partial^2 \omega}{\partial x^2}\right|_{x=0}
 = \left. \frac{\partial q}{\partial x}\right|_{x=0}
 \approx 4\frac{\omega(x_{+})-2\omega(x_{0})+\omega(x_{-})}{(x_+-x_-)^2},
\end{equation}
where $x_0=\tfrac{x_+ + x_-}{2}=0$.

When second derivative \eqref{eq:theory3} is small, results of the calculation
using \Eref{eq:theory2} are sufficiently reliable. If this were not the case
we would have strong interaction between levels. Such situation requires more
carefulness. It may be useful to trace the levels to smaller values of $x$.
Therefore, we do additional calculation for $x=-\tfrac38$. When theoretical
splitting for the interacting levels at $x=0$ is close to the experimental
one, then we can expect good accuracy for the $q$-factors even for this case.
Otherwise, if the splitting at $x=0$ differs from the experiment, the
$q$-factors calculated for $x=0$ may be incorrect. We can try to improve
calculated $q$-factors by moving along $x$ axis to the point where theoretical
splitting matches experiment \cite{DFK02}. As an additional test of the
accuracy of our theory we compare calculated $g$-factors with the experimental
ones from \cite{NIST}.

Absorption spectra in astrophysics correspond to the transitions from the
ground state, which in the case of Ni~II has $J=5/2$ and belongs to
configuration [Ar]$3d^9$. Since Ni~II has nine electrons in the open shells
its spectrum is dense and complicated \cite{NIST}. Due to the proximity of the
levels with the same total angular momentum $J$ and parity $P$  they may
strongly interact with each other, in particular for the high frequencies
which are more interesting for astrophysics. Because of the selection rules
for the transitions from the ground state we are interested in the states of
the negative parity with $J=3/2, 5/2, 7/2$.

In the frequency range observed in the high redshift astrophysics (i.e.\
roughly between 52000 and 70000 cm$^{-1}$) \cite{PTML00,RSGP12} there are five
lowest odd levels with $J=3/2$, eight levels with $J=5/2$, and seven levels
with $J=7/2$. All these levels belong to configurations $3d^{8}4p$,
$3d^{7}4s4p$, and $3d^{7}4p4d$. We performed calculations for all these
levels, as well as for the ground state for four values of $x$ and found
transition frequencies. Then we used cubic interpolation for the interval
$0.625<(x+1)< 1.275$.

We use Dirac-Coulomb Hamiltonian in the no-pair approximation. Breit
corrections to the $q$-factors were studied previously and were found to be
small \cite{PKR09}. We apply configuration interaction (CI) method in the
final basis set of relativistic orbitals. All calculations are done with the
computer package of I.~I.\ Tupitsyn \cite{BDT77,KT87}.

We start with solving the Hartree-Fock-Dirac equations. The self-consistency
procedure is done for the ground state configuration [Ar]$3d^{9}$. After that
all these shells are frozen. The valence orbital $4s$ is constructed for the
$3d^{8}4s$ configuration. Then two $4p_{j}$ orbitals are found for the
non-relativistic configuration $3d^{8}4p$. On the next stage we make virtual
orbitals using the method described in \cite{Bog91, KPF96}. In this method an
upper component of the virtual orbitals is formed from the previous orbital of
the same symmetry by multiplication with some smooth function of radial
variable $r$ in the spherical cavity with a radius 50 a.u. The lower component
is then formed using kinetic balance condition. This way we make finite basis
set of 33 orbitals that include $l=0,\ldots,3$ partial waves ($8spd6f$).

The configuration space is formed by making single and double (SD) excitations
from the small list of reference configurations. For even states this list
includes $3d^9$ and $3d^{8}4s$. For odd states the reference configurations
are: $3d^{8}4p$, $3d^{7}4s4p$, and $3d^{7}4p4d$. This way we get 3292 even and
2991 odd relativistic configurations. The number of configurations accounted
for in our present calculations is significantly bigger than in the previous
calculations in Ref.\ \cite{DFK02}.

\begin{table*}[htb]
\caption{Results of the calculations for Ni~II. Transition frequencies
$\omega$, $q$-factors, and $\partial q/ \partial x$  are given in \cm. Our
final recommended values with the errors in brackets are given in the column
$q_\mathrm{recom}$. In addition to the calculated $g$-factors we give their
values in the pure $LS$-coupling scheme. Experimental frequencies and
$g$-factors are taken from Ref.\ \cite{NIST}. Letters A --- H mark the levels
which interact in pairs (see discussion in the text). Asterisks mark the lines of prime astrophysical interest. }

\label{tbl:q-factors}
\begin{tabular}{lcldrddrrrr}
\hline \hline
& & \multicolumn{2}{c}{ Experiment} & \multicolumn{6}{c} {Theory} & \multicolumn{1}{c} {Ref.\cite{DFK02}} \\
\cline{5-10}
\multicolumn{1}{l}{} &    & \multicolumn{1}{c}{$\omega$}  & \multicolumn{1}{c}{$\quad g$} & \multicolumn{1}{c}{$\omega$} & \multicolumn{1}{c}{$\quad g_{calc}$} & \multicolumn{1}{c}{$\quad g(LS)$} & \multicolumn{1}{c}{$q$} & \multicolumn{1}{c}{$\partial q/ \partial x $} & \multicolumn{1}{c}{$q_\mathrm{recom}$} & \multicolumn{1}{c}{$q$} \\
\hline
${^4}D_{7/2}$ &    & 51558 & 1.420 & $\quad $49002 & 1.423 & 1.429 & $ \quad -2487$ $\quad$& $-159$ $\quad$& $-2490\,(150)$ & $-2415$ \\
${^4}D_{5/2}$ &    & 52739 & 1.365 & 50239 & 1.359 & 1.371 & $-1290$ $\quad$& $-240$ $\quad$& $-1290\,(150)$ & $-1231$ \\
${^4}D_{3/2}$ &    & 53635 & 1.186 & 51183 & 1.187 & 1.200 & $-313$  $\quad$& $-179$ $\quad$& $-310\,(150)$  &  \\
${^4}G_{7/2}$ &    & 54263 & 1.025 & 51693 & 1.010 & 0.984 & $-1393$ $\quad$& $-394$ $\quad$& $-1390\,(150)$ & $-1361$ \\
${^4}G_{5/2}$ &    & 55019 & 0.616 & 52482 & 0.609 & 0.571 & $-473$  $\quad$& $-359$ $\quad$& $-470\,(150)$  & $-394$   \\
${^4}F_{7/2}$ &    & 55418 & 1.184 & 53008 & 1.194 & 1.238 & $-1181$ $\quad$& $-211$ $\quad$& $-1180\,(150)$ & $-1114$ \\
${^4}F_{5/2}$ &    & 56075 & 0.985 & 53728 & 0.996 & 1.029 & $-409$  $\quad$& $-44$  $\quad$& $-410\,(150)$  & $-333$   \\
${^2}G_{7/2}$ &  A & 56371* & 0.940 & 53972 & 0.923 & 0.889 & $-134$  $\quad$& $-873$ $\quad$& $-250\,(300)$  & $-124$   \\
${^4}F_{3/2}$ &    & 56425 & 0.412 & 54140 & 0.420 & 0.400 & $-137$  $\quad$& $-370$ $\quad$& $-140\,(150)$  &  \\
${^2}F_{7/2}$ &  B & 57081* & 1.154 & 54817 & 1.134 & 1.143 & $-969$  $\quad$&$1358$  $\quad$& $-790\,(300)$  & $-700(250)$ \\
${^2}D_{5/2}$ &  C & 57420* & 1.116 & 55315 & 1.100 & 1.200 & $-1495$ $\quad$& $-200$ $\quad$& $-1500\,(150)$ & $-1400(250)$\\
${^2}F_{5/2}$ &  D & 58493* & 0.946 & 56376 & 0.966 & 0.857 & $-98$   $\quad$& $271$  $\quad$& $-100\,(150)$  & $-20(250)$ \\
${^2}D_{3/2}$ &    & 58706* & 0.795 & 56770 & 0.799 & 0.800 & $-367$  $\quad$& $-51$  $\quad$& $-370\,(150)$  &  \\
${^4}P_{5/2}$ &    & 66571* & 1.480 & 66169 & 1.506 & 1.600 & $-2205$ $\quad$& $-346$ $\quad$& $-2210\,(150)$ &  \\
${^4}P_{3/2}$ &  E & 66580 & 1.550 & 66173 & 1.592 & 1.733 & $-2286$ $\quad$& $-370$ $\quad$& $-2290\,(250)$ &  \\
${^2}F_{5/2}$ &    & 67695 & 0.960 & 67512 & 0.943 & 1.029 & $-1904$ $\quad$& $-161$ $\quad$& $-1900\,(150)$ &  \\
${^2}F_{7/2}$ &  G & 68131* & 1.200 & 67921 & 1.186 & 1.143 & $-1664$ $\quad$& $-376$ $\quad$& $-1600\,(200)$ &  \\
${^2}D_{3/2}$ &  F & 68154* & 1.020 & 68080 & 1.033 & 0.800 & $-1091$ $\quad$& $120$  $\quad$& $-1090\,(250)$ &  \\
${^2}D_{5/2}$ &    & 68736* & 1.264 & 68753 & 1.242 & 1.200 & $-410$  $\quad$& $-332$ $\quad$& $-410\,(150)$  &  \\
${^4}D_{7/2}$ &  H & 70778 & 1.385 & 70704 & 1.383 & 1.429 & $-662$  $\quad$& $530$  $\quad$& $-750\,(200)$  &  \\
\hline \hline
\end{tabular}
\end{table*}

\section{Results and Discussions}

\begin{figure*}[tbh]
\includegraphics[height=7.95cm, width=5.911cm]{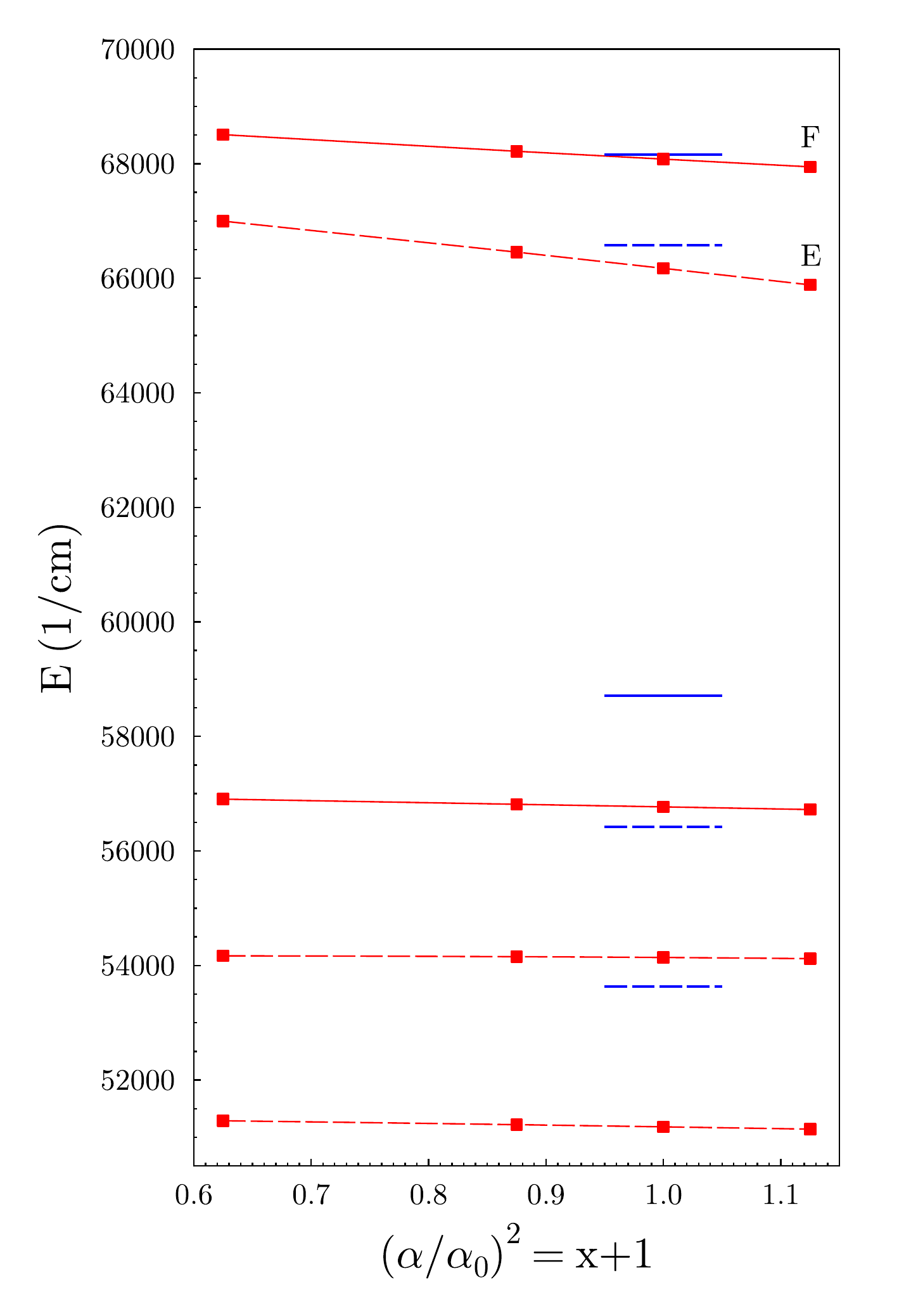}
\includegraphics[height=7.95cm, width=5.911cm]{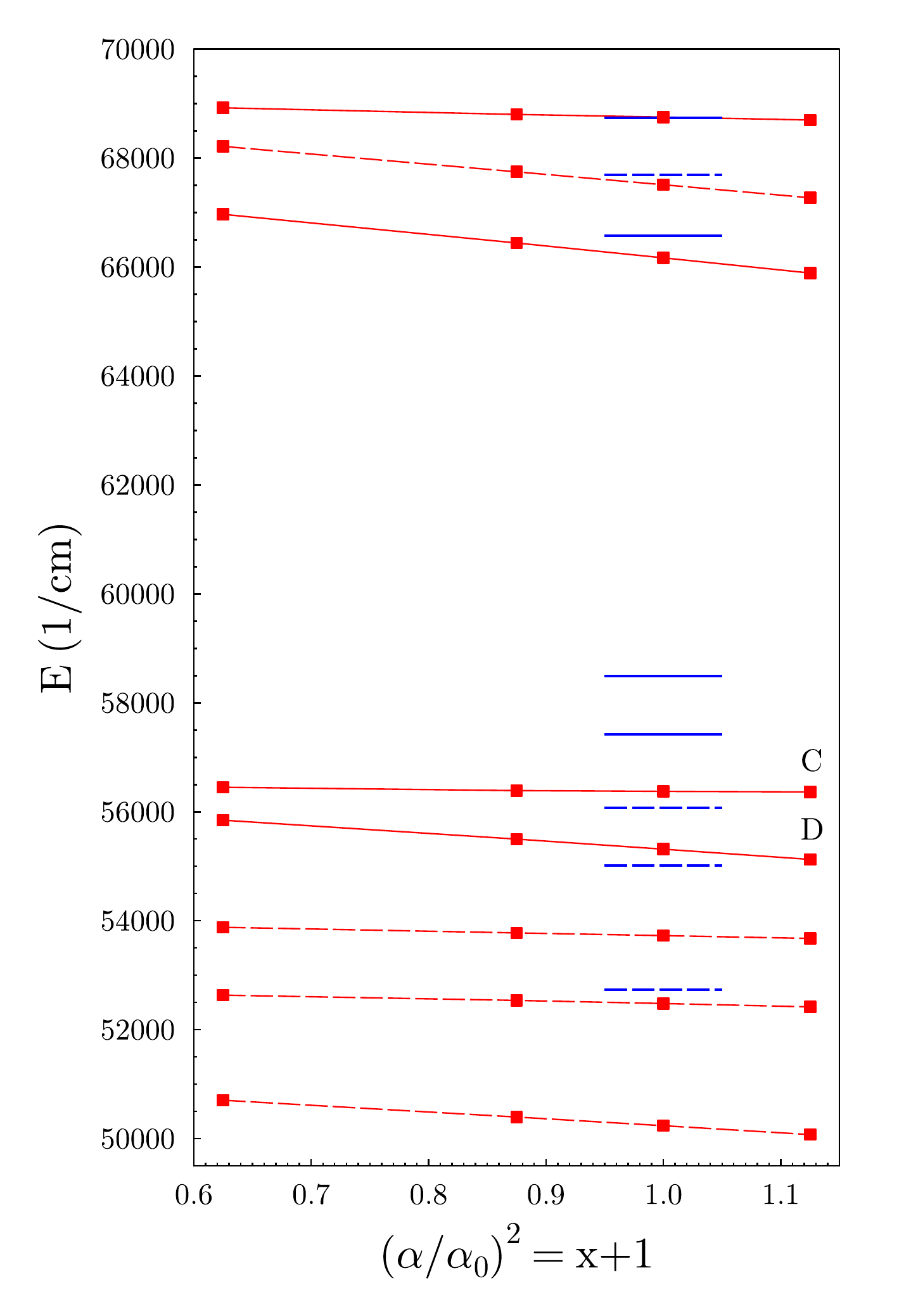}
\includegraphics[height=7.95cm, width=5.911cm]{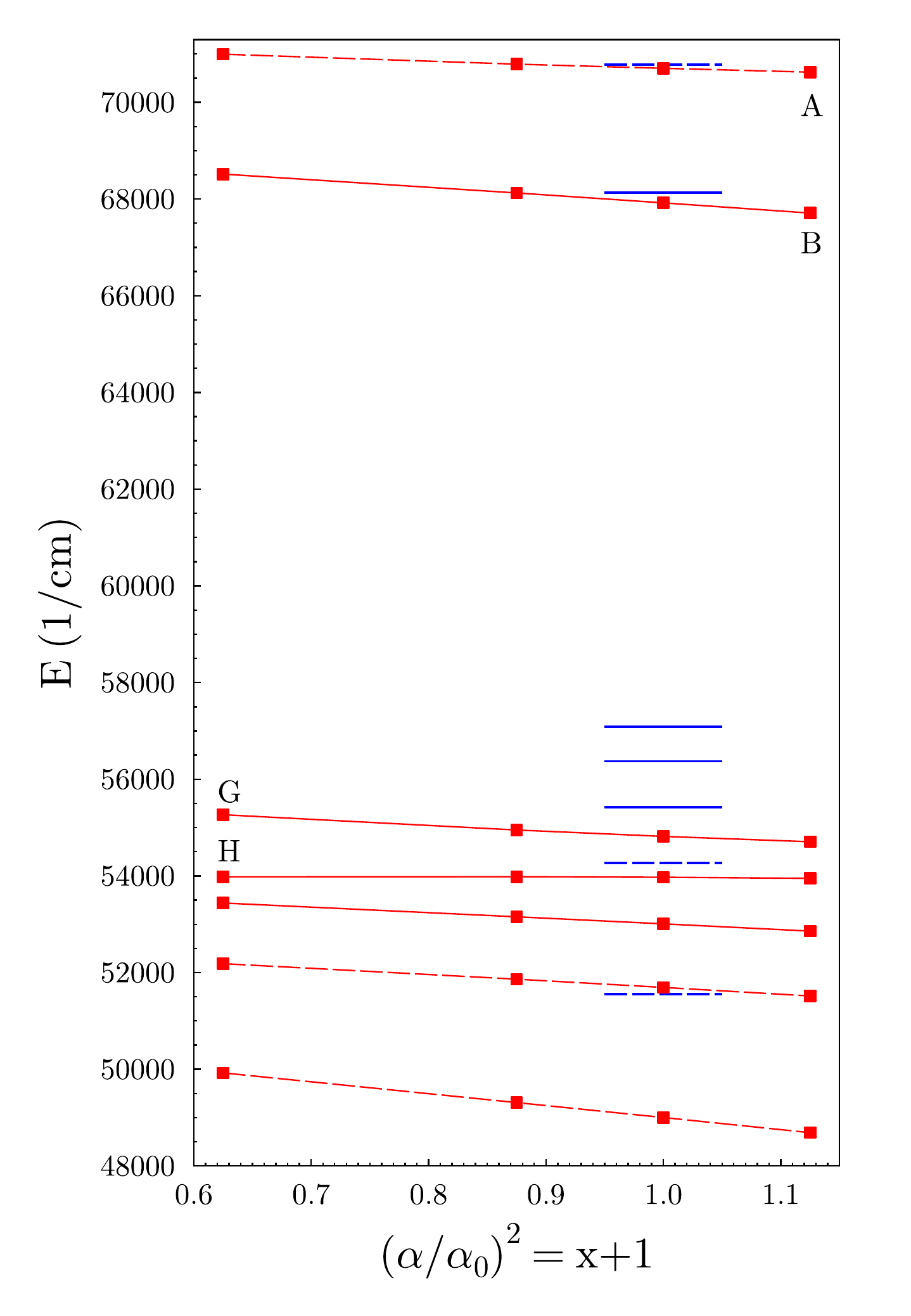}
\caption{The dependence of the transition frequencies from the ground state on
the parameter $(\alpha/\alpha_{0})^2 = x+1 $. Left panel corresponds to
$J=3/2$, central panel --- $J=5/2$, and right panel --- $J=7/2$. The solid
lines correspond to the transitions of astrophysical interest, dashed lines
--- to other transitions. The theoretical curves are obtained by cubic
interpolation between marked points. The short horizontal lines indicate
experimental data from Ref.\ \cite{NIST}. The $q$-factors correspond to the
slopes of the curves at $\alpha=\alpha_0$.}
 \label{alpha_dep}
\end{figure*}

Results of our calculations of the energies, $g$-factors, $q$-factors, and
their derivatives $\partial q/ \partial x$ for the odd levels are presented in
Table \ref{tbl:q-factors} and \fref{alpha_dep}. Calculations were done for
four values of $x$: $x = -\tfrac38, -\tfrac18,\, 0, +\tfrac18$. Then we did
cubic interpolation for the interval $-\tfrac38\le x\le +\tfrac18$. The
$q$-factors were found by using \Eref{eq:theory2} and by differentiation of
the interpolation polynomial at $x=0$. Both results appeared to be very close.

Identification of the levels is done by comparison of the calculated
$g$-factors with the experimental ones and with the prediction of the
$LS$-coupling scheme (see \tref{tbl:q-factors}). For the energy range in
question the order of the calculated levels agrees with the experiment.
Moreover, one can see from the figure that our calculation in general
reproduces the splittings between the nearest levels, but somewhat
underestimates the energies of the lower levels of negative parity.

For most levels the derivative $\partial q/\partial x$ is rather small and
negative. That means that interaction between neighboring levels with the same
quantum numbers $J^P$ is not very strong. However there are four levels in
\tref{tbl:q-factors} with $\partial q/\partial x >0$. In each of these cases
there is close level with large negative $\partial q/\partial x$. Therefore,
we can assume that these levels are at pseudo crossing and there is strong
repulsion between them. Such interacting pairs of levels are marked on
\fref{alpha_dep}. Below we analyze each of these pairs separately.

\paragraph*{Levels A$({}^2G_{7/2})$ and B$({}^2F_{7/2})$.}

This is the strongest interacting pair of levels with $\tfrac{\partial
q_A}{\partial x}$=$-873\,$\cm\ and  $\tfrac{\partial q_B}{\partial
x}$=$1358\,$\cm. Calculated energy splitting, 845 \cm, is larger than
experimental splitting 709 \cm. It is clear from \fref{alpha_dep} that pseudo
crossing takes place at positive values of $x$. Both calculated $g$-factors
differ from the experiment by roughly 2\%. If we move along $x$ axis towards
the crossing point at $x=x^*\approx 0.28$ we get smaller splitting and can
expect better agreement with the experiment. Indeed, the splitting for
$x=\tfrac18$ is 758 \cm, which is almost three times closer to the
experimental value. The error for $g$-factors also decreases to  1\%, or so.
We conclude that this way we do get better agreement with the experiment.

At the pseudo crossing $x=x^*$ the levels are parallel to each other, so
$q_A^*=q_B^*$. Therefore, the shift towards $x^*$ strongly affects calculated
$q$-factors of each level leaving their sum almost constant. At $x=\tfrac18$
we get $q_A = -255$ \cm\ and $q_B=-792$ \cm. To get exactly experimental
splitting we need to move even closer to $x^*$. However, the interaction of the
levels A and B with other levels is not at all negligible and we need to
reproduce other energy splittings as well. This can not be done with the help
of the single parameter $x$. Therefore, we take $q$ factors at $x=\tfrac18$ as
our recommended values. Theoretical error here is, of course, quite large. We
estimate it to be around 300 \cm, which covers the whole range of $x$ between
$x=0$ and $x=x^*$.

\paragraph*{Levels C$({}^2D_{5/2})$ and D$({}^2F_{5/2})$.}

These levels interact much weaker than the first pair with $\tfrac{\partial
q_C}{\partial x}$=$-200\,$\cm\ and $\tfrac{\partial q_D}{\partial
x}$=$271\,$\cm. Besides, the calculated energy splitting, 1061 \cm, is very
close to the experimental splitting 1073 \cm. One of the two $g$-factors is
2\% larger than experimental value, but another one is much closer. We
conclude that no correction is necessary for this pair of levels.

\paragraph*{Levels E$({}^4P_{3/2})$ and F$({}^2D_{3/2})$.}

Though these levels have derivatives of the opposite sign, their absolute
values differ by a factor of 3. That means that the two level model is not
applicable here. The energy splitting is overestimated by 332 \cm, or about
20\% and $g$-factor of the level E is 3\% larger than experimental value.
Because of the strong interaction of this pair with other levels we do not
introduce any correction, but rather use the shift in $x$ to estimate the
error. Experimental splitting is reproduced at $x=-0.3$ where $q$-factors
appears to be: $q_E=-2140$ \cm\ and $q_F=-1180$ \cm. Therefore, we estimate
the error to be about 250 \cm\ for both levels.

\paragraph*{Levels G$({}^2F_{7/2})$ and H$({}^4D_{7/2})$.}

Interaction of these two levels is a little stronger than for the pair (C, D)
and theoretical energy splitting, 2647 \cm\ is  135 \cm\ smaller, than
required. Calculated $g$-factors are quite good, but the difference between
them is small and we can not use them to estimate the mixing of these levels
with each other. The optimal splitting corresponds to $x=-0.14$ where $q_G =
-1600$ \cm\ and $q_H=-750$ \cm. We estimate the error for the $q$-factors to
be about 200 \cm.

For the remaining levels, which do not form strongly interacting pairs, the
slope remains relatively constant. Calculated $q$-factors are, therefore, less
sensitive to the details of the calculation. We checked that any shifts along
$x$ axis in order to improve energy splittings between neighboring levels do
not change $q$-factors by more than 100 \cm. Thus, we estimate theoretical
error for these $q$-factors to be around 150 \cm. Our final recommended values
for the $q$-factors with the error bars are listed in \tref{tbl:q-factors}.
Approximately half of the levels from this table were studied before in Ref.\
\cite{DFK02}. For all of them we have good agreement between two calculations.
At the same time both calculations do not agree with the earlier calculation
\cite{MWFD01}.

\begin{table}[h]
\caption{ Oscillator strengths $f_\mathrm{osc}$ for the transitions considered
in this paper. Calculations were done in the length (L) and velocity (V) gauges.
Were possible we compare our results with compilation of \citet{Mor03} and
with Ref.\ \cite{HS13}.} \label{tbl:fosc}
\begin{tabular}{llcccc}
\hline \hline
& \multicolumn{1}{c}{$\omega$}
& \multicolumn{4}{c} {$f_\mathrm{osc}$} \\
\cline{3-6}
& \multicolumn{1}{c}{\cm}
& \multicolumn{1}{c}{$\,$ L-gauge$\,$}
& \multicolumn{1}{c}{$\,$ V-gauge$\,$}
& \multicolumn{1}{c}{$\,$ Ref.\ \cite{HS13}$\,$}
& \multicolumn{1}{c}{$\,$ Ref.\ \cite{Mor03}$\,$}\\
\hline
${^4}D_{7/2}$ & 51558  &  2.99E-08 & 9.83E-11& \\
${^4}D_{5/2}$ & 52739  & 5.86E-04 & 5.72E-04 & \\
${^4}D_{3/2}$ & 53635  & 1.35E-05 & 1.50E-05 & \\
${^4}G_{7/2}$ & 54263  & 3.63E-04 & 3.30E-04 & \\
${^4}G_{5/2}$ & 55019  & 6.99E-05 & 7.33E-05 & \\
${^4}F_{7/2}$ & 55418  & 3.55E-03 & 3.14E-03 & 7.16E-03 \\
${^4}F_{5/2}$ & 56075  & 5.39E-04 & 5.33E-04 & \\
${^2}G_{7/2}$ & 56371* & 2.22E-03 & 1.95E-03 & 6.22E-03 \\
${^4}F_{3/2}$ & 56425  & 6.13E-05 & 6.92E-05 & \\
${^2}F_{7/2}$ & 57081* & 2.75E-02 & 2.53E-02 & 2.77E-02 & 2.77E-02 \\
${^2}D_{5/2}$ & 57420* & 5.05E-02 & 5.03E-02 & 4.27E-02 & 4.27E-02 \\
${^2}F_{5/2}$ & 58493* & 4.50E-02 & 4.36E-02 & 3.24E-02 & 3.24E-02 \\
${^2}D_{3/2}$ & 58706* & 1.04E-02 & 1.06E-02 & 6.00E-03 & 6.00E-03 \\
${^4}P_{5/2}$ & 66571* & 5.24E-03 & 4.76E-03 & 6.00E-03 \\
${^4}P_{3/2}$ & 66580  & 4.40E-04 & 3.12E-04 \\
${^2}F_{5/2}$ & 67695  & 6.94E-04 & 8.64E-04 &          & 9.72E-04 \\
${^2}F_{7/2}$ & 68131* & 1.02E-02 & 1.20E-02 & 9.90E-03 & 9.90E-03 \\
${^2}D_{3/2}$ & 68154* & 8.87E-03 & 8.03E-03 & 6.30E-03 & 6.30E-03 \\
${^2}D_{5/2}$ & 68736* & 3.03E-02 & 2.87E-02 & 2.76E-02 & 3.23E-02 \\
${^4}D_{7/2}$ & 70778  & 3.15E-03 & 3.59E-03 \\
\hline \hline
\end{tabular}
\end{table}

\paragraph*{Oscillator strengths.}

Observability of the transitions considered here depends on the respective
oscillator strengths $f_\mathrm{osc}$. Not all of them are known from the
experiment. We calculated E1 transition amplitudes and $f_\mathrm{osc}$ in the
length (L) and velocity (V) gauges for all transitions from
\tref{tbl:q-factors}. Results are summarized in \tref{tbl:fosc}. For the
transitions with $f_\mathrm{osc}>10^{-3}$ two calculations agree within 10\%.
Even for most weaker transitions $10^{-3}>f_\mathrm{osc}>10^{-4}$ the
agreement is quite good. Only in one case the difference is close to 30\%.
Usually, for the CI wave functions the difference between amplitudes in L- and
V-gauges can be used to estimate the accuracy of the theory. Therefore we
expect our results for the transitions with $f_\mathrm{osc}>10^{-4}$ to be
accurate within 20-30\%. Within this error bar our results agree with
compilation of experimental and observational results by \citet{Mor03}. The
only exception is the line 58706 \cm, where our strength is 50\% larger than
Morton's.

We also have mostly good agreement with Ref.\ \cite{HS13}. For the line 68736
\cm\ our values lie between those of Refs.\ \cite{Mor03} and \cite{HS13}.
However, for the lines 55418 and 56371 \cm\ our results are significantly
smaller. In particular, for the potentially interesting line 56371 \cm\ our
strength is three times smaller than in \cite{HS13}. This makes observation of
this line for the high redshift sources more difficult.

\section{Conclusion}

To summarize, here we present theoretical $q$-factors for Ni~II found with CI
method for Dirac-Coulomb Hamiltonian in no-pair approximation. We calculated
$q$-factors for several new lines, which had not been studied theoretically
before, but were observed in the high redshift quasar spectra. All calculated
sensitivities for Ni~II are negative. Two new lines have relatively small
$q$-factors, around $-400$ \cm\ and one has $q$-factor, which is one of the
largest in absolute value, $q=-2210$ \cm. Note, that large difference in
sensitivities of individual lines increases sensitivity of the observations to
$\alpha$-variation and allows effective control of the systematics.

In comparison to the previous calculations we have significantly increased the
CI space. That did not changed results very much and we have good agreement
with the most resent calculation \cite{DFK02}. We do not think the accuracy
can be noticeably improved within CI method: Ni~II has 9 electrons in open
shells and it is practically impossible to saturate CI space. On the other
hand, already present accuracy is sufficient to analyze astrophysical data on
the possible $\alpha$-variation.

\acknowledgments We thank H. Rahmani and R. Srianand for stimulating this
research and J. Berengut and M. Murphy for interesting discussions. This work
is partly supported by the Russian Foundation for Basic Research Grant No.\
14-02-00241.

\end{document}